# $(Li_{0.84}Fe_{0.16})OHFe_{0.98}Se$ superconductor: Ion-exchange synthesis of large single crystal and highly two-dimensional electron properties

Xiaoli Dong[1*†], Kui Jin[1*], Dongna Yuan[1*], Huaxue Zhou[1*§], Jie Yuan[1], Yulong Huang[1], Wei Hua[2], Junliang Sun[2], Ping Zheng[1], Wei Hu[1], Yiyuan Mao[1], Mingwei Ma[1], Guangming Zhang[3], Fang Zhou[1†], and Zhongxian Zhao[1†]

[1]Beijing National Laboratory for Condensed Matter Physics, Institute of Physics, Chinese Academy of Science, Beijing 100190, China
[2]Beijing National Laboratory for Molecular Sciences, College of Chemistry and Molecular Engineering, Peking University, Beijing 100871, China
[3]State Key Laboratory of Low-Dimensional Quantum Physics and Department of Physics, Tsinghua University, Beijing 100084, China

**Abstract:**
A large and high-quality single crystal $(Li_{0.84}Fe_{0.16})OHFe_{0.98}Se$, the optimal superconductor of newly reported $(Li_{1-x}Fe_x)OHFe_{1-y}Se$ system, has been successfully synthesized via a hydrothermal ion-exchange technique. The superconducting transition temperature ($T_c$) of 42 K is determined by magnetic susceptibility and electric resistivity measurements, and the zero-temperature upper critical magnetic fields are evaluated as 79 and 313 Tesla for the field along the *c*-axis and the *ab*-plane, respectively. The ratio of out-of-plane to in-plane electric resistivity, $\rho_c/\rho_{ab}$, is found to increases with decreasing temperature and to reach a high value of 2500 at 50 K, with an evident kink occurring at a characteristic temperature $T^*=120$ K. The negative in-plane Hall coefficient indicates that electron carriers dominate in the charge transport, and the hole contribution is significantly reduced as the temperature is lowered to approach $T^*$. From $T^*$ down to $T_c$, we observe the linear temperature dependences of the in-plane electric resistivity and the magnetic susceptibility for the FeSe layers. Our findings thus reveal that the normal state of $(Li_{0.84}Fe_{0.16})OHFe_{0.98}Se$ becomes highly two-dimensional and anomalous prior to the superconducting transition, providing a new insight into the mechanism of high-$T_c$ superconductivity.

The binary $Fe_{1+\delta}Se$ compound [1, 2] structured with FeSe-tetrahedron layers of a maximum inter-layer compactness exhibits a superconductivity only at 8.5 K under ambient pressure. However, in a highly two-dimensional single unit-cell FeSe film on a $SrTiO_3$ substrate, the superconducting transition temperature $T_c$ can be increased up to a value as high as 65 K [3-6]. In other more complicated FeSe-based superconductor compounds, where the FeSe-tetrahedron layers with a larger inter-layer spacing serve all the same as basic superconducting units, the superconductivity of a higher $T_c$ than binary bulk $Fe_{1+\delta}Se$ is always realized. In $A_yFe_{2-x}Se_2$ compounds, for example, the alkali metal ions A are intercalated into between the adjacent FeSe-tetrahedron layers, expanding the inter-layer separation and yielding superconductivity with $T_c \sim 30$ K. It seems that the superconductivity is related with the structural and/or electronic two-dimensionality to some extent. However, the superconducting phase in $A_yFe_{2-x}Se_2$ is always intergrown with an insulating antiferromagnetic (AFM) phase resulting from a $\sqrt{5}\times\sqrt{5}$ superstructure of Fe vacancies [7]. This complexity obstructs further experimental studies on its intrinsic properties of both the normal and superconducting states.

Recently, $(Li_{1-x}Fe_x)OHFe_{1-y}Se$ (FeSe11111) superconductors with $T_c$ up to 40 ~ 43 K have been synthesized by hydrothermal approaches [8-11]. The troublesome $\sqrt{5}\times\sqrt{5}$ Fe vacancy ordered phase is *absent* [11], but a canted AFM order in the intercalated (Li/Fe)OH layers is reported to coexist with the superconductivity [8]. Since FeSe11111 has a larger inter-layer spacing than that of $A_yFe_{2-x}Se_2$, the correlation between the adjacent FeSe layers becomes even weaker. Such a quasi two-dimensional structure is propitious for unveiling the interplay of electronic anisotropy and high-$T_c$ superconductivity in the FeSe-based superconductors. For in-depth investigations on the intrinsic electronic properties of FeSe11111, high-quality large single crystals are indispensable.

In this Letter, we report the successful synthesis of a large and high-quality single crystal $(Li_{0.84}Fe_{0.16})OHFe_{0.98}Se$ via a hydrothermal ion-exchange technique using big insulating $K_{0.8}Fe_{1.6}Se_2$ crystal as a matrix. Its superconducting transition is confirmed by magnetic susceptibility and electric resistivity measurements, and the zero-temperature upper critical magnetic fields are estimated to be 79 and 313 Tesla for the field along the *c*-axis and the *ab*-plane, respectively. The temperature dependences of out-of-plane and in-plane electric resistivity are measured and their ratio, $\rho_c/\rho_{ab}$, as a measure of the charge transport anisotropy, is found to increase with decreasing temperature and to reach its large value of 2500 at 50 K, showing an evident kink at a characteristic temperature $T^*$=120 K. Meanwhile, the negative in-plane Hall coefficient indicates that electron carriers dominate the charge transport and the hole contribution is significantly reduced as the temperature is lowered to approach $T^*$. From $T^*$ down to $T_c$, there emerge the linearly temperature dependent in-plane electric resistivity and the magnetic susceptibility deduced for the FeSe layers. Hence, our results reveal that the electron carriers of $(Li_{0.84}Fe_{0.16})OHFe_{0.98}Se$ superconductor are getting highly two-dimensional prior to the superconducting transition, resulting in the anomalous normal state properties.

The current hydrothermal methods using commercially available reagents together with, in a particular case, synthesized binary FeSe [10] as starting materials have only produced powder samples of FeSe11111 [8-11]. Our recipe for preparing the large and high quality crystals is that large crystals of $K_{0.8}Fe_{1.6}Se_2$ (nominal 245 phase, KFS245 for short) are specially grown and used as a kind of *matrix* for a hydrothermal ionic exchange reaction. The KFS245 structure consists of an alternative stacking with layers of K ions and FeSe-tetrahedra similar to our target compound, so as to be the best precursor with an ideal matrix structure. Moreover, the K ions in KFS245 are *completely* released into solution after the hydrothermal reaction process. Simultaneously, (Li/Fe)OH layers constructed by ions from the solution are squashed into the matrix, linking the adjacent edge-sharing FeSe-tetrahedra via a weak hydrogen bonding. The separation between the neighboring FeSe-tertahedron layers is consequently enlarged by ~32% (from ~ 7.067 Å to ~ 9.318 Å), resulting in an enhanced two-dimensionality of the electronic structure compared with both the bulk FeSe and the intercalated $A_yFe_{2-x}Se_2$ superconductors.

As-derived high-quality FeSe11111 single crystals are of a size over 10 mm in length and about 0.4 mm in thickness. The cartoon in Figure 1a illustrates such an ionic exchange process. The experimental details of crystal synthesis are described in Supplemental Material. The insets of Fig. 1b and c are the photographs of two typical crystal pieces for KFS245 precursor and its subsequently derived FeSe11111, respectively. Interestingly, the FeSe11111 crystal roughly inherits the original shape of its precursor. The room temperature X-ray diffraction (XRD) patterns of (00*l*) type on the two crystal pieces demonstrate their crystal orientations along (001) planes, with *l* = 2*n* (*n* = integer) for KFS245 (Fig.1b) and without the systematic extinction for FeSe11111 (Fig.1c). Their corresponding powder XRD patterns are given in the Supplemental Materials Fig.S1a and S1b, showing no detectable impurity phases within experimental resolution. All the reflections in each powder pattern can be well indexed on known tetragonal structures with the space group of *I*4/m [7] for KFS245 and *P*4/nmm [8-10] for FeSe11111. The least-squares refined unit cell dimensions are *a* = 8.7248 (6) Å and *c* = 14.1339 (12) Å for KFS245, and *a* = 3.7827 (4) Å and *c* = 9.3184 (7) Å for FeSe11111. Single crystal XRD data of FeSe11111 are also collected at 180 K and the structure refinement is performed in light of the reported structure model [8, 10]. The resulting crystallographic data are listed in the Supplemental Materials Table S1. Finally the chemical formula of our single crystal is expressed as $(Li_{0.84}Fe_{0.16})OHFe_{0.98}Se$ based on the structural refinement. While the √5×√5 ordered vacant Fe sites in the FeSe layers of $K_{0.8}Fe_{1.6}Se_2$ are of ~20% in amount [7], most of them are occupied by Fe ions imported during the ionic exchange process, leaving over there only 2% disordered Fe vacancies in the end-crystal of FeSe11111. The atomic ratio of Li:Fe:Se (0.838:1.142:1, Supplemental Materials Table S1) determined from the structural refinement is in perfect agreement with the result (0.82:1.17:1) from inductively coupled plasma atomic emission spectroscopy (ICP-AES) analysis. Importantly, no trace of K ions in our FeSe11111 crystals is detectable by energy dispersive X-ray spectrometry (EDX) and ICP-AES, verifying a complete release of K ions after the hydrothermal ionic exchange.

Fig. 2a displays the magnetic susceptibility data for the KFS245 precursor, exhibiting an AFM phase transition at 538 K consistent with the previous reports [7]. Its insulating behavior is checked up by electric resistivity measurement (Supplemental Materials Fig. S2). By contrast, the ion-exchange synthesized FeSe11111 crystal exhibits bulk superconductivity, as evidenced by its magnetic susceptibilities drop at 42 K, with a narrow temperature range $\Delta T \sim 1$ K between 10% and 90% shielding signals. The sharp diamagnetic transition along with its 100% diamagnetic shielding demonstrates a high quality of our FeSe11111 single crystal.

Displayed in Fig. 3a and 3b are the temperature dependences of the in-plane electric resistivity $\rho_{ab}(T)$ under the magnetic fields along the *c*-axis (*H//c*) and the *ab*-plane (*H//ab*), respectively. The onset superconducting transition temperature remains almost the same under fields up to 9 Tesla, no matter *H//c* or *H//ab*. However, the zero resistance state shifts to lower temperatures with increasing magnetic field, but more slowly in *H//ab* than in *H//c*, implying the formation of complex vortex states. Such a behavior resembles the features observed in copper-oxide high temperature superconductors [12, 13]. By taking the temperature at the mid-point of the electric resistivity drop as that of the upper critical field $H_{c2}$ (Fig.3c), the zero-temperature upper critical magnetic fields can be estimated as 79 T for field along the *c*-axis and 313 T for field along the *ab*-plane from the Werthamer-Helfand-Hohenberg (WHH) formula [14] (inset of Fig.3c). So the superconducting state of the FeSe11111 sample is anisotropic.

In the normal state, the in-plane electric resistivity $\rho_{ab}$ displays a metallic behavior in the whole measuring temperature range (Fig. 4a). As the temperature is lowered to the characteristic temperature $T^*=120$ K, the slope of the resistivity curve undergoes a prominent reduction. Subsequently, there emerges the linear temperature dependence of $\rho_{ab}$ from 80 K down to $T_c$ (inset of Fig. 4a). Due to the large size of our sample, the out-of-plane electric resistivity of FeSe11111 single crystal is readily measured for the first time. The resulting $\rho_c$ shows a roughly similar behavior to $\rho_{ab}$. However, their ratio $\rho_c/\rho_{ab}$, as a measure of the carrier anisotropy, increases with decreasing temperature and reaches to a large value of 2500 at 50 K (Fig.4a). Interestingly, a kink in the ratio also appears at the characteristic temperature $T^*$, due to the significant slope change in $\rho_{ab}$ compared to $\rho_c$ (Supplemental Materials Fig. S3). It is clear that, prior to the superconducting transition, the normal state electronic properties of the FeSe11111 crystal turns out to be highly two-dimensional.

Actually, other normal state properties are also readily measured on our large FeSe11111 crystal. The in-plane Hall voltage has been obtained under the magnetic fields along the *c*-axis by sweeping the field up to 9 Tesla at fixed temperatures. The extracted Hall resistivity ($\rho_{xy}$) is proportional to the magnetic field at all measuring temperatures (Supplemental Material Fig. S4), so the Hall coefficient $R_H = \rho_{xy}/H$ is field independent. As shown in Fig. 4b, $R_H$ is negative in the whole measuring temperature, but displays a dip feature around the characteristic temperature $T^* = 120$ K. We propose to use a compensated metal picture of multiple electron and hole bands to explain our observation. With the equal total electron and hole concentrations ($n_h = n_e$), the Hall coefficient satisfies $R_H = (\mu_h-\mu_e)/[(\mu_h+\mu_e)n_e e]$, where $\mu_e$ ($\mu_h$) stands for the mobility of electron (hole) charge carriers. The negative Hall coefficient indicates that the contribution of electrons dominates the charge transport of the sample, $\mu_e > \mu_h$. Above the characteristic temperature $T^*$, the mobility of holes $\mu_h$ is reduced as the temperature is getting lower, leading to a decreasing Hall coefficient. Below $T^*$, the electron carriers govern the charge transport, $\mu_e \gg \mu_h$. The Hall coefficient can thus be simplified as $R_H \sim -1/(n_e e)$, and its upturn with the further lowering temperature can be attributed to an enhanced carrier density.

The static magnetic susceptibility $\chi = M/H$ is carefully measured and analyzed under different magnetic fields $H$ along $c$-axis in the region with linear field dependences of magnetization $M$ (Supplemental Material Fig. S5). There exist two different contributions to the magnetic susceptibility: one part arises from the intercalated (Li/Fe)OH layers and another part from the FeSe layers. The total susceptibility slightly depends on the magnitude of the applied field (Fig. 4c). In the higher temperature range, a modified Curie-Weiss law is obeyed: $\chi_m = \chi_0 + \chi_{CW}$, where $\chi_{CW} = C/(T-\theta)$. The effective magnetic moment in the (Li/Fe)OH layers is fitted as $\mu_{Fe} \approx 4.2 \sim 4.6\ \mu_B$, consistent with the previous results [10]. These local magnetic moments are nearly free. But the fitted $\theta$ is of a small negative value (-6.6 K ~ -7.5 K), implying an AFM long-range order will eventually form in the (Li/Fe)OH layers in the lower temperature regime [8]. A constant term $\chi_0$ signifies a Pauli paramagnetic contribution from itinerant charge carriers. But a deviation from the modified Curie-Weiss law is evident again below the characteristic temperature $T^*$ (=120 K). When the Curie-Weiss term $\chi_{CW}$ is subtracted, the rest part of magnetic susceptibility depends almost linearly on temperature in the range of 55 K ~ 75 K (inset of Fig. 4c), suggesting the presence of two-dimensional AFM spin fluctuations in the FeSe layers. Such a linear behavior has been previously seen in the paramagnetic metallic states of iron arsenide superconductors [15, 16]. Therefore, the prominent slope reduction in the $\rho_{ab}$-$T$ curve around $T^*$ is likely due to the additional scattering mechanism brought about by the emergent two-dimensional AFM spin fluctuations and the vanishing mobility of holes below $T^*$.

Finally, we have established for the first time a common temperature scale ($T^*$=120 K) in $(Li_{0.84}Fe_{0.16})OHFe_{0.98}Se$ superconductor, which characterizes the prominent reduction in the slope of the in-plane electric resistivity, the kink in the ratio $\rho_c/\rho_{ab}$, the dip in the in-plane Hall coefficient, and the deviation in the magnetic susceptibility from the Curie-Weiss law. Below this characteristic $T^*$, normal state electronic behaviors are getting highly two-dimensional and the AFM spin fluctuations in the FeSe layers set in, manifested by the anomalous linear in-plane electric resistivity and nearly linear magnetic susceptibility deduced for the FeSe layers. The two-dimensional electron-electron interaction is crucial to the electron pairing of superconductivity in the FeSe layers, as commonly believed in optimal iron arsenide [17-19] and copper-oxide high temperature superconductors [20-22], as well as in heavy fermion [23, 24] and organic superconductors [25]. Hence our results suggest that the $(Li_{0.84}Fe_{0.16})OHFe_{0.98}Se$ superconductor falls under the same universality class.

In conclusion, we have successfully synthesized large and high quality superconducting $(Li_{0.84}Fe_{0.16})OHFe_{0.98}Se$ single crystal out of insulating $K_{0.8}Fe_{1.6}Se_2$ matrix crystal via hydrothermal ion exchange. Its in-plane lattice constants $a$ and $b$ determined as 3.78 Å are comparable to the primary bulk FeSe (3.77 Å) and $A_yFe_{2-x}Se_2$ (3.90 Å) superconductors. In contrast, its FeSe-tetrahedron layers weakly bonded to the (Li/Fe)OH layers via hydrogen ions have a larger inter-layer spacing (9.32 Å) compared with the bulk FeSe (5.52 Å) and $A_xFe_{2-y}Se_2$ (7.02 Å) superconductors. Consequently, $(Li_{0.84}Fe_{0.16})OHFe_{0.98}Se$ exhibits an enhanced two-dimensionality of the electronic structure among iron selenide superconductors, prior to the superconducting transition at 42 K. Hence our findings provide a new insight into the pairing mechanism of high-$T_c$ superconductivity.

Furthermore, a series of large single crystals of $(Li_{1-x}Fe_x)OHFe_{1-y}Se$ system are expected by duplicating the hydrothermal ion-exchange technique. The breakthrough in crystal preparation reported here will greatly promote the in-depth investigations on this system. In addition, this very successful technique developed for synthesizing $(Li_{0.84}Fe_{0.16})OHFe_{0.98}Se$ crystals may be applicable in other layer-structured materials, opening up a new route for producing and exploring high-quality bulk functional crystals desired for both basic research and applications.

**Acknowledgements** This work is supported by the National Basic Research Program of China (projects 2013CB921701), "Strategic Priority Research Program (B)" of the Chinese Academy of Sciences (No. XDB07020100) and Natural Science Foundation of China (projects 11274358, 11190020 and 91222107).

* First equal authors
† Corresponding authors: dong@iphy.ac.cn; fzhou@iphy.ac.cn; zhxzhao@iphy.ac.cn
§ Permanent address: College of Physics, Chongqing University, Chongqing 401331, China

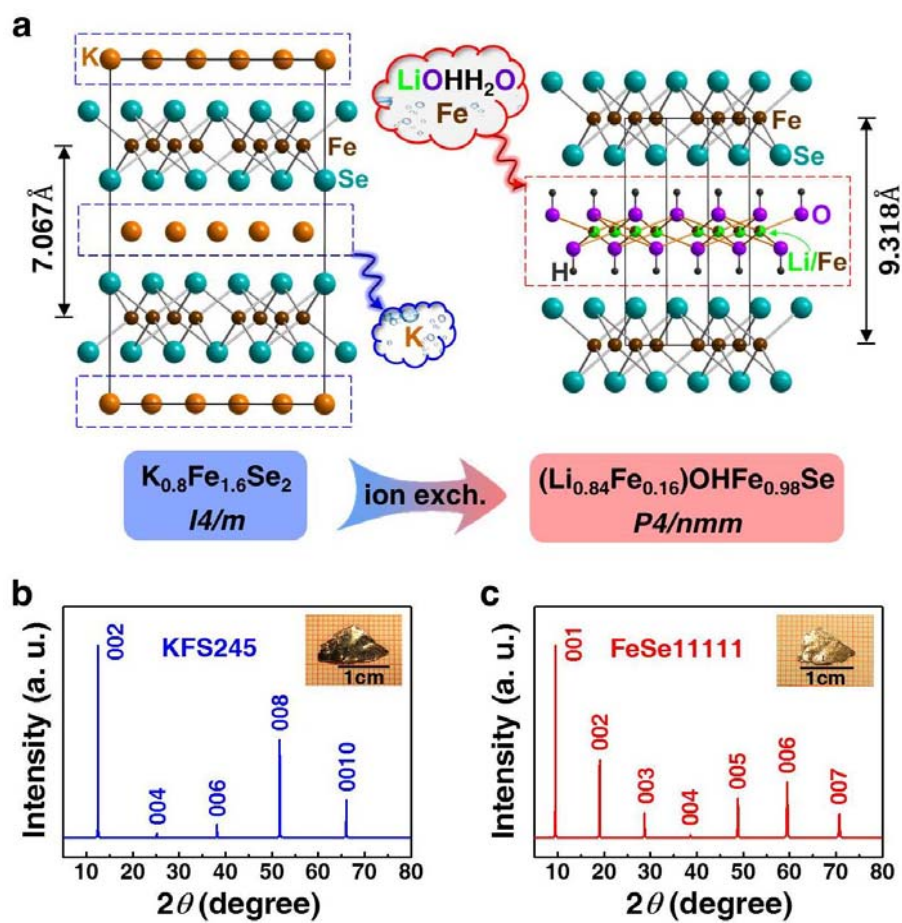

**Figure 1 Illustration of hydrothermal ion-exchange process.** **(a)** A schematic illustration of the hydrothermal ionic exchange reaction with the starting materials of big matrix crystal of $K_{0.8}Fe_{1.6}Se_2$, $LiOH \cdot H_2O$, Fe and $CH_4N_2Se$. The spacing between the adjacent FeSe layers of thus synthesized $(Li_{0.84}Fe_{0.16})OHFe_{0.98}Se$ crystal with the space group $P4/nmm$ is consequently enlarged by ~ 32 % (from ~ 7.067 Å in $K_{0.8}Fe_{1.6}Se_2$ to ~ 9.318 Å in $(Li_{0.84}Fe_{0.16})OHFe_{0.98}Se$). For clarity, the H-Se and K-Se bondings in the structures are not shown. **(b)** and **(c)** are the XRD patterns of (00*l*) type for the $K_{0.8}Fe_{1.6}Se_2$ and the $(Li_{0.84}Fe_{0.16})OHFe_{0.98}Se$ crystal pieces, respectively, demonstrating their crystal orientations along (001) planes. The insets show the corresponding photographs of the crystals.

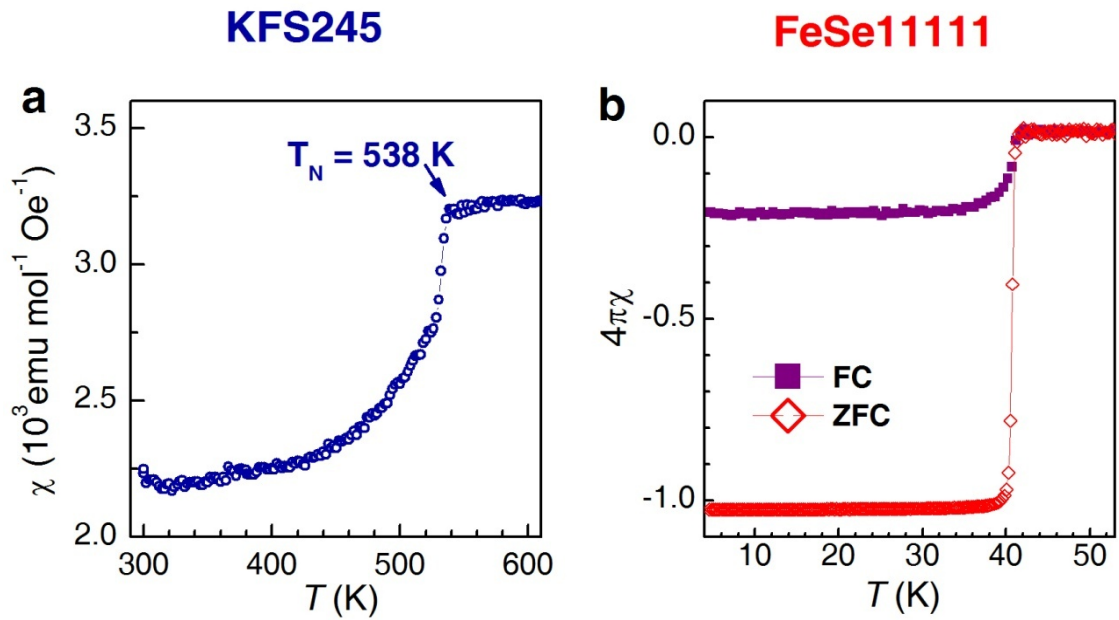

**Figure 2 Magnetic properties of $K_{0.8}Fe_{1.6}Se_2$ and $(Li_{0.84}Fe_{0.16})OHFe_{0.98}Se$ crystals.**
**(a)** The temperature dependence of the static magnetic susceptibility of $K_{0.8}Fe_{1.6}Se_2$ exhibits an antiferromagnetic transition at 538 K. **(b)** The magnetic susceptibilities corrected for demagnetization factor under zero-field cooling (ZFC) and field cooling (FC, 1 Oe along *c*-axis) for $(Li_{0.84}Fe_{0.16})OHFe_{0.98}Se$ shows that a sharp diamagnetic transition occurs at 42 K. The ZFC curve shows a 100 % superconducting shielding and the FC one a Meissner signal up to ~21%.

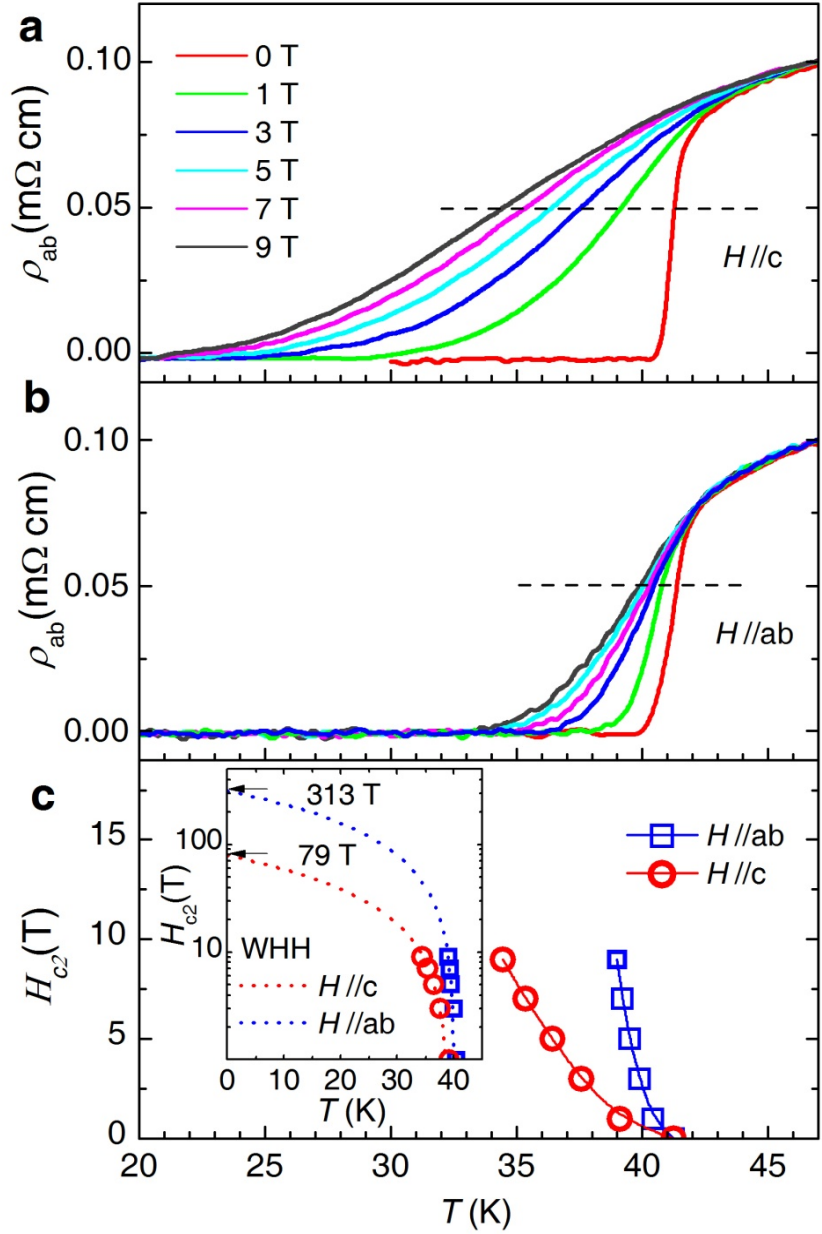

**Figure 3 The in-plane electric resistivity under different magnetic fields and the upper critical magnetic fields.** (a) The in-plane electric resistivity of $(Li_{0.84}Fe_{0.16})OHFe_{0.98}Se$ displays the superconducting transitions under different applied magnetic fields along $c$-axis. (b) The in-plane electric resistivity under different applied magnetic fields along $ab$-plane. Up to 9 Tesla, the onset superconducting transition temperature remains almost unchanged in both (a) and (b), but the temperature of zero resistance state shifts quickly from 40 K to lower temperatures with increasing field. (c) The anisotropic upper critical magnetic field $H_{c2}$ as functions of temperature deduced from (a) and (b), from which the zero-temperature upper critical fields can be extrapolated according to the Werthamer-Helfand-Hohenberg (WHH) formula (the inset).

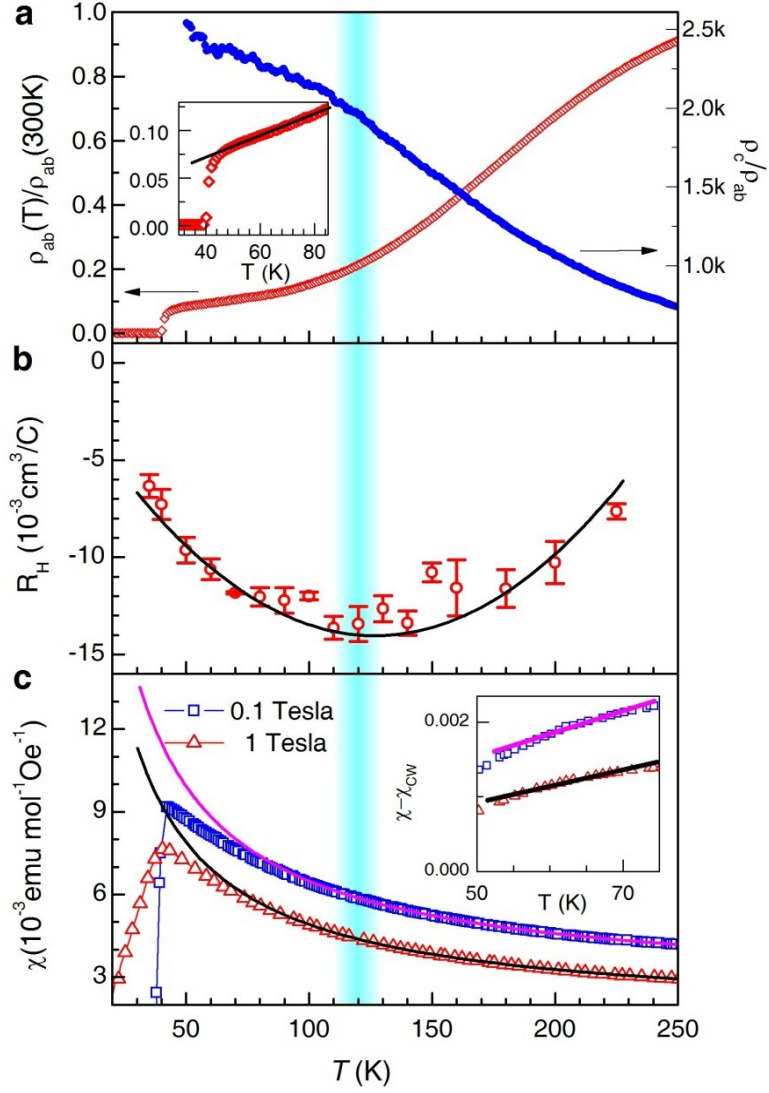

**Figure 4 The electric resistivity, Hall coefficient, and static magnetic susceptibility of $(Li_{0.84}Fe_{0.16})OHFe_{0.98}Se$ single crystal.** (a) The in-plane electric resistivity and the ratio of the out-of-plane resistivity to in-plane one as functions of temperature. Around $T^*=120$ K, the slope of the resistivity curve undergoes a prominent reduction and a corresponding kink occurs in the curve of $\rho_c/\rho_{ab}$. The inset figure shows the linearly temperature dependent range from 80 K down to $T_c$ for $\rho_{ab}$. (b) The in-plane Hall coefficient $R_H$ as a function of temperature exhibits a dip feature around $T^*=120$ K. The black curve is guide to the eye. (c) The temperature dependences of static magnetic susceptibility $\chi = M/H$ under magnetic fields along c-axis. The results *slightly* depend on the magnitude of the applied field. The sudden drops in the magnetic susceptibility are due to the appearance of the superconductivity. In the high temperature range, all the data can be fitted to a modified Curie-Weiss law $\chi_m = \chi_0 + \chi_{CW}$. But a deviation is clearly visible below $T^*=120$ K as well. Further lowering temperature, the rest magnetic susceptibility in which the Curie-Weiss term is subtracted shows the nearly linear temperature dependence in a narrow temperature range (the inset).